\documentclass[%
 reprint, superscriptaddress,noeprint,nolongbibliography,
 aps,
]{revtex4-2}

\usepackage{graphicx}% Include figure files
\usepackage{dcolumn}% Align table columns on decimal point
\usepackage{bm}% bold math
\usepackage{times}
\usepackage{xr}
\usepackage{graphicx}
\usepackage{bm}
\usepackage{amsmath}
\usepackage{amssymb}
\usepackage[dvipsnames]{xcolor}
\usepackage{hyperref}
\usepackage{siunitx,xfrac}
\usepackage{xcolor}

\usepackage{etoolbox}
\usepackage{nomencl}
\makenomenclature

\usepackage[left=0.5in, right=0.5in, top=0.5in, bottom=1in]{geometry}

\begin{document}

\preprint{APS/123-QED}

\title{The physics of crêpes: Elasto-gravity control of soft folding}

\author{
Tom Marzin$^{1,2}$, Barath Venkateswaran$^{1}$, Yuchen Xi$^{1}$, Sunghwan Jung$^{2}$and P.-T. Brun$^{3}$.
\\
\small{$^{1}$Department of Chemical and Biological Engineering,}
\small{Princeton University, Princeton, NJ 08540, USA}\\
\small{$^{2}$ Department of Biological and Environmental
Engineering,}
\small{Cornell University, Ithaca, NY 14853, USA}\\
\small{$^{3}$Department of Chemical Engineering, Soft Matter, Rheology and Technology (SMaRT), KU Leuven, Leuven, Belgium}
}

\date{\today}% It is always \today, today,

\begin{abstract}
Like a crêpe resting on a plate, a thin elastic sheet can fold smoothly under its own weight, forming reversible shapes without creases or imposed hinges. Such soft folds arise from a balance between elastic bending and gravity, yet their stability, packing limits, and dynamics remain poorly understood. Here we show that these behaviors are governed by a single physical length scale, the elasto-gravity length $\ell_{eg}$. Using experiments and heavy-elastica theory, we demonstrate that $\ell_{eg}$ sets the characteristic fold geometry, determines when a fold becomes unstable and unfolds, and limits how many reversible folds can be stacked in rectangular and circular sheets. In particular, when lengths are rescaled by $\ell_{eg}$, fold shapes and stability thresholds collapse across materials and thicknesses. We further show that unfolding follows a universal speed scaling $v \sim \sqrt{g\,\ell_{eg}}$, revealing a gravity-controlled time scale for the release of stored bending energy. Together, these results establish a unified physical framework for reversible folding, compact storage, and gravity-assisted deployment of thin elastic sheets.
\end{abstract}

\maketitle

\begin{figure}[h!]
    \centering
    \includegraphics[width=\columnwidth]{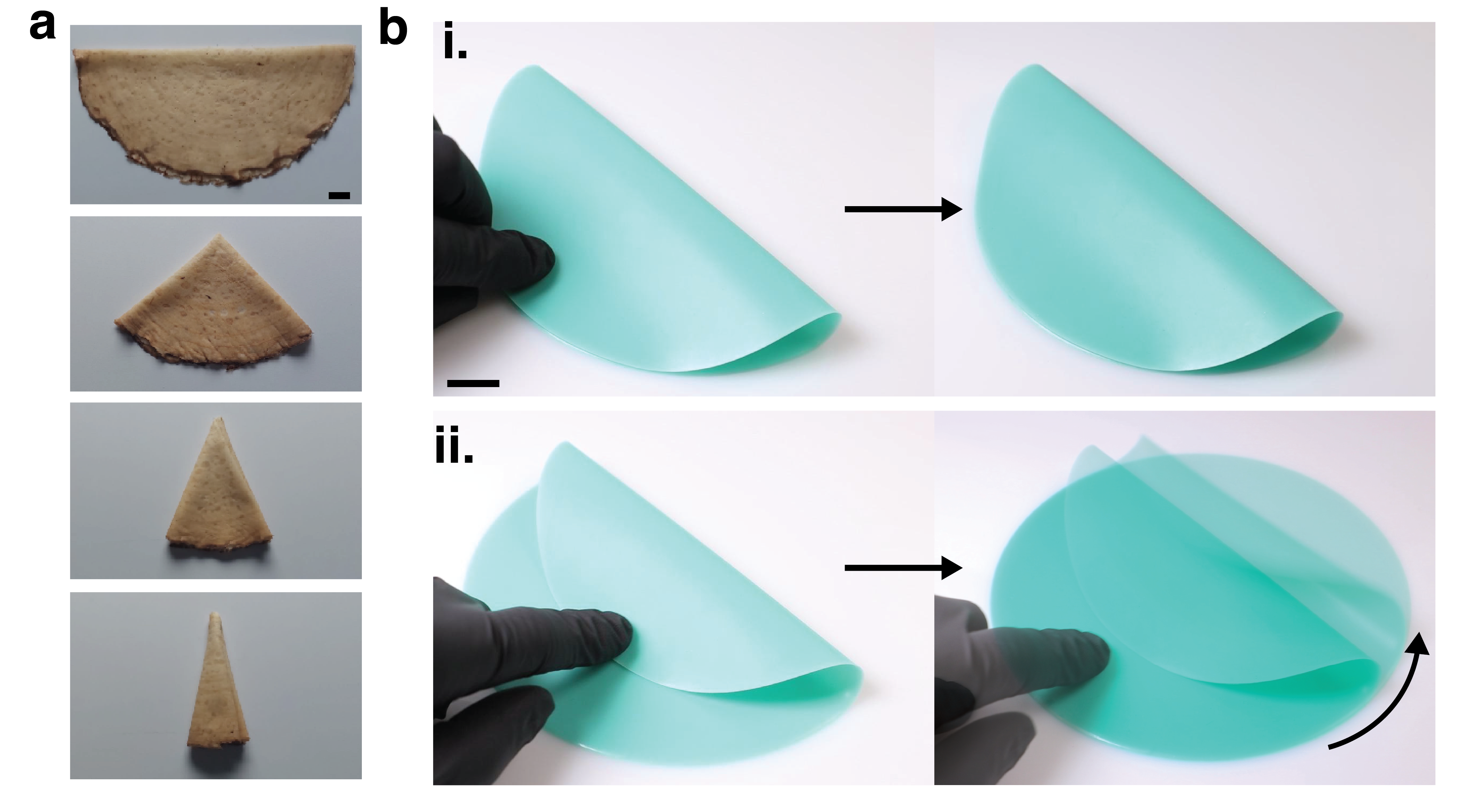} \caption{\textbf{Soft folding:} a) Multi folded shape of a crêpe of radius $R \approx 13~\text{cm}$ and $\ell_{eg} \approx 0.86~\text{cm}$. b) Elastomeric circular membrane of radius $R = 7.5~\text{cm}$ and $\ell_{eg} = 1.99~\text{cm} $ wrapped in two configurations: i) a symmetric half-fold that forms a stable loop, ii) an asymmetric, partial fold that is mechanically unstable and unravels upon release. (scale bars are $2$ cm)}
    \label{Fig:1}
    %(Young’s modulus $E \approx 884~\text{kPa}$ \Tom{TC})
\end{figure}

Folding a sheet in half is a familiar operation, encountered in activities ranging from everyday handling to industrial packaging. In food preparation, for example, wraps such as tortillas or crêpes are routinely folded (Fig.~\ref{Fig:1}a), while in manufacturing, thin sheets are folded or rolled to increase storage density and later enable deployment. In some contexts, folding is achieved by imposing sharp creases~\cite{jules2019local} or by applying external forces that fix the fold shape~\cite{paulsen2015optimal,py2007capillary}. By contrast, many thin sheets can also fold softly under their own weight, forming smooth and reversible configurations without creases (Fig.~\ref{Fig:1}b). Whether such folds remain stable, or relax and reorganize, is governed not by yield or plasticity but by a subtle, nonlinear interplay among geometry, elastic bending, and gravity. Such folding behavior is particularly relevant in cases where large, thin sheets must be stored compactly without creases, thereby avoiding persistent damage from plasticity effects.

%Understanding these criteria is especially useful for storing large, thin sheets without persistent crease damage due to plasticity effects. In this Letter, we will predict when a folded sheet will persist, collapse, or spontaneously unfold.

In many problems involving thin structures, the material's weight is assumed to be negligible. This is the case for inflated membranes, where tension and pressure dominate ~\cite{merritt1978pressure,vella2011wrinkling}, or for snapping instabilities in which elastic bending alone is key to elucidate the stability~\cite{gomez2017critical,marzin2025augmented}. Here, we instead consider heavy elastic sheets for which gravity directly competes with bending elasticity ~\cite{wang_critical_1986,batista2020mechanics}. 
%This elasto-gravitational balance gives rise to smooth, reversible folded configurations whose shapes are selected by out-of-plane bending under self-weight. 
Related competitions between bending and gravity have been explored in a variety of systems, including rucks in rugs ~\cite{vella_statics_2009,kolinski2009shape}, gravity-induced buckling ~\cite{taffetani2019limitations,baroudi_nonlinear_2019,tani2024curvy}, and wrinkle or fold formation under confinement ~\cite{brau_wrinkle_2013,deboeuf2024yin,alben2022packing,davidovitch2021rucks}. However, in contrast to origami-like creases ~\cite{lechenault2014mechanical,jules2019local}, compression-induced folds ~\cite{marzin2025augmented}, or naturally curved slender systems ~\cite{callan-jones_self-similar_2012,arriagada2014curling}, where deformation is either localized or geometrically prescribed, folds in our system are not imposed. Instead, they emerge, migrate, and disappear through the interplay of bending, gravity, and frictional contact forces.
As a first step towards understanding these temporary folds, we will predict whether a folded sheet will persist or spontaneously unfold.

In this Letter, we combine experiments and theory to characterize the static and dynamic behavior of soft folded states (Fig.~\ref{Fig:1}b). In these configurations, part of the sheet rests flat on a rigid substrate, while the remaining portion lifts and bends out of plane before reconnecting with the flat region, forming a smooth and reversible fold. This soft folding mechanism is observed both in controlled polymer and plastic sheets and in familiar materials, including food samples such as crêpes and tortillas, for which multiple folds can be formed and stacked without creasing or permanent deformation (Fig.~\ref{Fig:1}a). Here we show how the competition between bending stiffness and self-weight selects a characteristic fold size and determines whether a given folded state is stable or instead spontaneously unravels (Fig.~\ref{Fig:1}b and Movie~S1).

Beyond identifying stability criteria, we translate our results into practical design principles, e.g.,
%. We first show how the position and size of a fold control a sheet’s propensity to unwrap, yielding quantitative guidelines
for stacking multiple folds without triggering spontaneous release. We then demonstrate that, once stability is lost, unfolding occurs through a rapid conversion of bending and gravitational energy into motion, with characteristic speeds set by the elasto-gravity balance and the sheet's material properties. Together, these results provide a unified, mechanics-based framework for understanding soft folding in heavy elastic sheets. By connecting the behavior of familiar folded objects to underlying elastic principles, they offer both intuitive, pedagogical insight and practical design rules for gravity-mediated storage and deployment.

\begin{figure}[h!]
    \includegraphics[width=1\columnwidth]{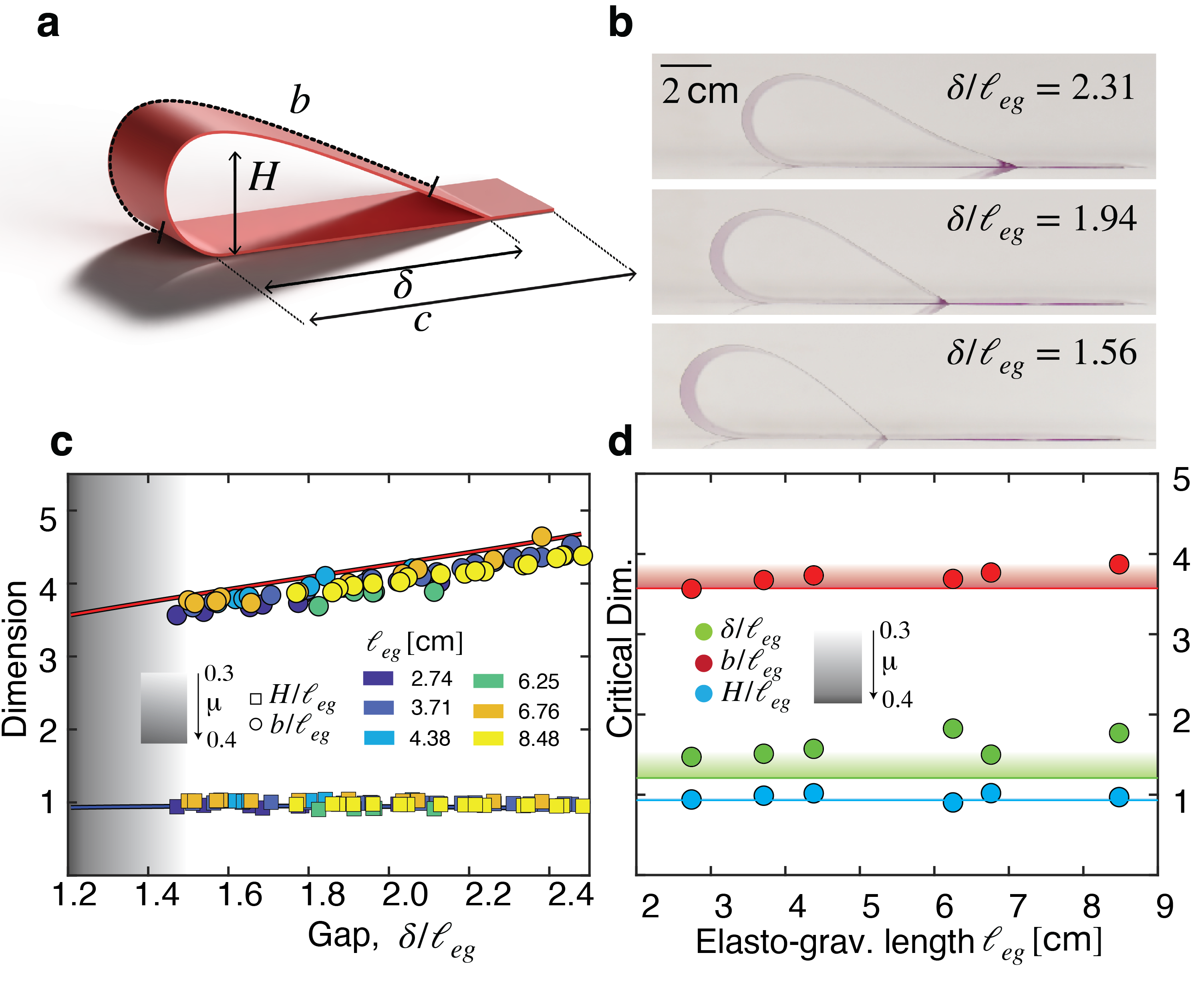}
    \caption{\textbf{Folding stability:} a) Schematic of the soft folding geometry. 
    b) Three representative folding configurations at different gap values $\delta$, for a strip $L=30$ cm long and $t=40$ $\mu$m thick ($\ell_{eg}= 3.71 ~\text{cm}$). 
    c) Fold length and height as functions of the gap for various sheet thicknesses; all lengths are rescaled by the elasto-gravity length $\ell_{eg}$.  
    d) Dimensions of the fold at criticality (length, height, and gap) as functions of $\ell_{eg}$. In c) and d), solid lines represent theoretical predictions, while shaded regions indicate the range expected from frictional effects. }
  %Shaded areas in (c) and (d) show the influence of the friction coefficient $\mu$.
    \label{Fig:2}
\end{figure}

Fig.\ref{Fig:2} reports on the stability of folded configurations with slender elastic sheets (Shimstock Plastic Sheets, Artus) of thicknesses $h \in [25,125]\mathrm{\mu m}$, width $w = 6\,\text{cm}$, lengths $L\in\{30,50,60\}\,\mathrm{cm}$, and Young's modulus $E = 4.76 \pm 0.46\,\mathrm{GPa}$. One end of each sheet is positioned onto a substrate. The other end is brought into contact with the flat portion of the sheet to create a soft \textit{fold} (Fig.~\ref{Fig:2}a-b), which is a smooth, reversible, arch-like configuration formed without creasing or plastic deformation. The primary control parameter is the end-to-end gap $\delta$ between the two contact points of the sheet, which is adjusted quasi-statically by gently sliding the upper layer (Fig.~\ref{Fig:2}a--b). For each imposed value of $\delta$, we measure the fold height $H$ and the free length $b$, defined as the portion of the sheet not in contact with the substrate (Fig.\ref{Fig:2}c). An experiment terminates when the folded configuration loses stability and spontaneously unravels to the flat state. We denote this instability threshold by $\delta_-$ and record the corresponding values $H_-$ and $b_-$. Note that when the gap is sufficiently large, $\delta>\delta_+$, the end portion of the sheet lies flat on the lower one, producing a finite dead length, similar to the elastocapillary racket~\cite{py2007capillary}. These configurations are, by definition, stable. 

We focus on the interval $\delta \in [\delta_-, \delta_+]$, for which a stable, well-defined fold without excess flat length exists. By balancing elastic and gravitational moment, we obtain an elasto-gravity length scale~\cite{vella_statics_2009,wang_critical_1986,holmes2019elasticity},
% \textcolor{red}{SJ: You didn't use $w$ here. I guess $b$ up there is $w$? } 
\begin{equation}
\ell_{eg}=\left(\frac{B}{\rho gw h}\right)^{1/3},
\label{Eq.elasto-gravity}
\end{equation}
where $B=Eh^3w/12(1-\nu^2)$ is the sheet's bending stiffness, $\rho$ the density of the sheet, and $g$ the acceleration due to gravity. Considering an infinitesimal deformation $\epsilon$ and balancing the gravitational and bending moments, we have $\rho g h w \ell_{eg}\epsilon \sim B\epsilon/\ell_{eg}^2$, thereby recovering Eq.~\ref{Eq.elasto-gravity}. Given the nature of our problem, we anticipate the relevance of $\ell_{eg}$, which varies between $2.7$ and $8.5,\mathrm{cm}$ in our experiments (see SI). In Fig.~\ref{Fig:2}c, we plot $H/\ell_{eg}$ and $b/\ell_{eg}$ versus $\delta/\ell_{eg}$. Our experimental data collapses across sheet thicknesses, indicating that fold shapes are indeed governed by $\ell_{eg}$. In particular, $H\approx\ell_{eg}$ is a very good estimate for the fold height, regardless of the gap, a remarkably simple result.

In Fig.~\ref{Fig:2}d, we report the dimensions of a fold near threshold. Remarkably, these values are nearly constant when rescaled by $\ell_{eg}$.  To further rationalize this observation, we model the free portion of the sheet as an inextensible \emph{heavy} elastica. The sheet is uniform along its width $w$, and deformation is confined to the $x$–$y$ plane. The centerline is parameterized by a dimensionless arc-length $s\in[0,1]$ rescaled by the \textit{a priori}, unknown free length $b$.  The coordinates are given by  $x'=\cos\theta$ and $y'=\sin\theta$, where and $\theta(s)$ is the angle made by the tangent to the centerline with the horizontal. Force resultants are written in dimensionless form as $(n_x,n_y)$, where $n_x$ is the unknown horizontal force and $n_y'=(b/\ell_{eg})^{3}$ is the gravity component. Moment balance yields the second order heavy-elastica equation ~\cite{wang_critical_1986,atanackovic1997stability,vella_statics_2009} (see SI for further details):
\begin{equation}
\theta'' + n_y \cos\theta - n_x \sin\theta \;=\; 0.
\label{Eq.elastica}
\end{equation}
Closing the problem requires seven boundary conditions. In the absence of adhesion, the curvature vanishes at both contact lines, reflecting the absence of localized moments at these points~\cite{vella_statics_2009,kolinski2009shape}. We thus have $x(0)=0$, $y(0)=0$, $\theta(0)=0$ and $\theta'(0)=0$, at the base $s=0$; and $y(1)=0$ and $\theta'(1)=0$ at the free end $s=1$. The end-to-end gap sets the final condition $x(1)=\delta$. As $\delta$ increases, loss of contact occurs when the vertical force at the upper contact vanishes, triggering unfolding. The model predicts this instability at $\delta_-/\ell_{eg}=1.21$, with corresponding dimensions $b_-/\ell_{eg}=3.57$ and $H_-/\ell_{eg}=0.93$ (Fig.~\ref{Fig:2}d).

% For each $\delta$, we solve the boundary-value problem and extract the corresponding values of $H$ and $b$. As $\delta$ increases, the vertical contact force $n_y$ at $s=1$ flips its sign from positive to negative. In the absence of any adhesive force, the sheet can no longer remain in contact with itself, and hence unfolds for $n_y=0$. The model predicts unfolding happens at a critical rescaled gap $\delta_-/\ell_{eg}= 1.21$, with corresponding fold dimensions $b_-/\ell_{eg}=\,3.57$ and $H_-/\ell_{eg}=\, 0.93$ (horizontal lines in Fig.~\ref{Fig:2}d). 

% For each $\delta$, we solve the boundary-value problem to obtain $H$ and $b$. As $\delta$ increases, the vertical contact force $n_y$ at $s=1$ changes sign. n the absence of adhesion, loss of contact and unfolding occur at $n_y=0$. The model predicts a critical gap $\delta_-/\ell_{eg}=1.21$, with corresponding dimensions $b_-/\ell_{eg}=3.57$ and $H_-/\ell_{eg}=0.93$ (Fig.~\ref{Fig:2}d).

In Fig.~\ref{Fig:2}c-d, we show that the shapes obtained by integrating Eq.~\ref{Eq.elastica} are mostly in favorable agreement with experiments. However, we observe a small yet systematic deviation in our estimates of $b$ and $\delta$. In effect, experiments suggest that the transition occurs at $\delta_-^{exp}/\ell_{eg} \approx 1.61 \pm 0.15$, which corresponds to  $b_-^{exp}/\ell_{eg} \approx 3.71 \pm 0.1$ (see Fig.\ref{Fig:2}d). This discrepancy suggests the presence of an additional instability mechanism. Thus, we impose another stability criterion, based on frictional forces acting on the free end. Using Coulomb friction, a surface with static friction coefficient $\mu$ can support horizontal forces at the sheet's free end without slipping, up to a certain threshold, i.e., $|n_x| < \mu |n_y|$. Including this constraint in our model, we find that low static friction coefficients lead to a premature loss of stability, as the upper layer slips over the lower layer before lifting off. The corresponding friction-limited region is shown by the shaded bands in Fig.~\ref{Fig:2}d for $0.3 \lesssim \mu \lesssim 0.4$. Experiments indicate a transition around $\mu \approx \, 0.28 \pm 0.04$, consistent with reported friction values for similar materials~\cite{yoshida2020mechanics,sano2023randomly}. Notice that for sufficiently large friction ($\mu > 0.39$), the model predicts that the sheet can only undergo a flip instability rather than slipping.

We now turn to large values of delta, beyond which both extremities of the loop lie flat on the substrate. This critical gap is obtained by integrating Eq.~\ref{Eq.elastica}. We find $\delta_+/\ell_{eg}=2.48$, while experiments yield a slightly larger value, $\delta_+^{exp}/\ell_{eg}=2.63 \pm 0.16$. The corresponding free lengths are $b_+/\ell_{eg}=4.68$ in theory and $b_+^{exp}/\ell_{eg}=4.69 \pm 0.29$ experimentally (see SI for experimental points), showing good agreement in this regime.

Beyond stability, the geometry of a soft fold determines its capacity to enclose material.  We quantify this capacity through the volume enclosed by the folded configuration. Measurements and predictions from Eq.~\ref{Eq.elastica} show good agreement across all sheet thicknesses (see graph and derivation details in SI). Notably, the enclosed volume varies non-monotonically with the imposed gap $\delta$ and reaches a maximum at an optimal shift $\delta_*=2.02\,\ell_{eg}$. At this optimum, the volume reaches $1.51 \ell_{eg}^2 w$, irrespective of the material thickness. Experiments yield $\delta_*^{\mathrm{exp}}=(2.18 \pm 0.15) \ell_{eg}$ and a corresponding volume $(1.62 \pm 0.13)\ell_{eg}^2 w$, confirming the existence of an optimal fold that maximizes enclosed capacity while remaining mechanically stable. Next, we examine the stability of stacked soft folds.

\begin{figure}[h!]
    \includegraphics[width=1\columnwidth]{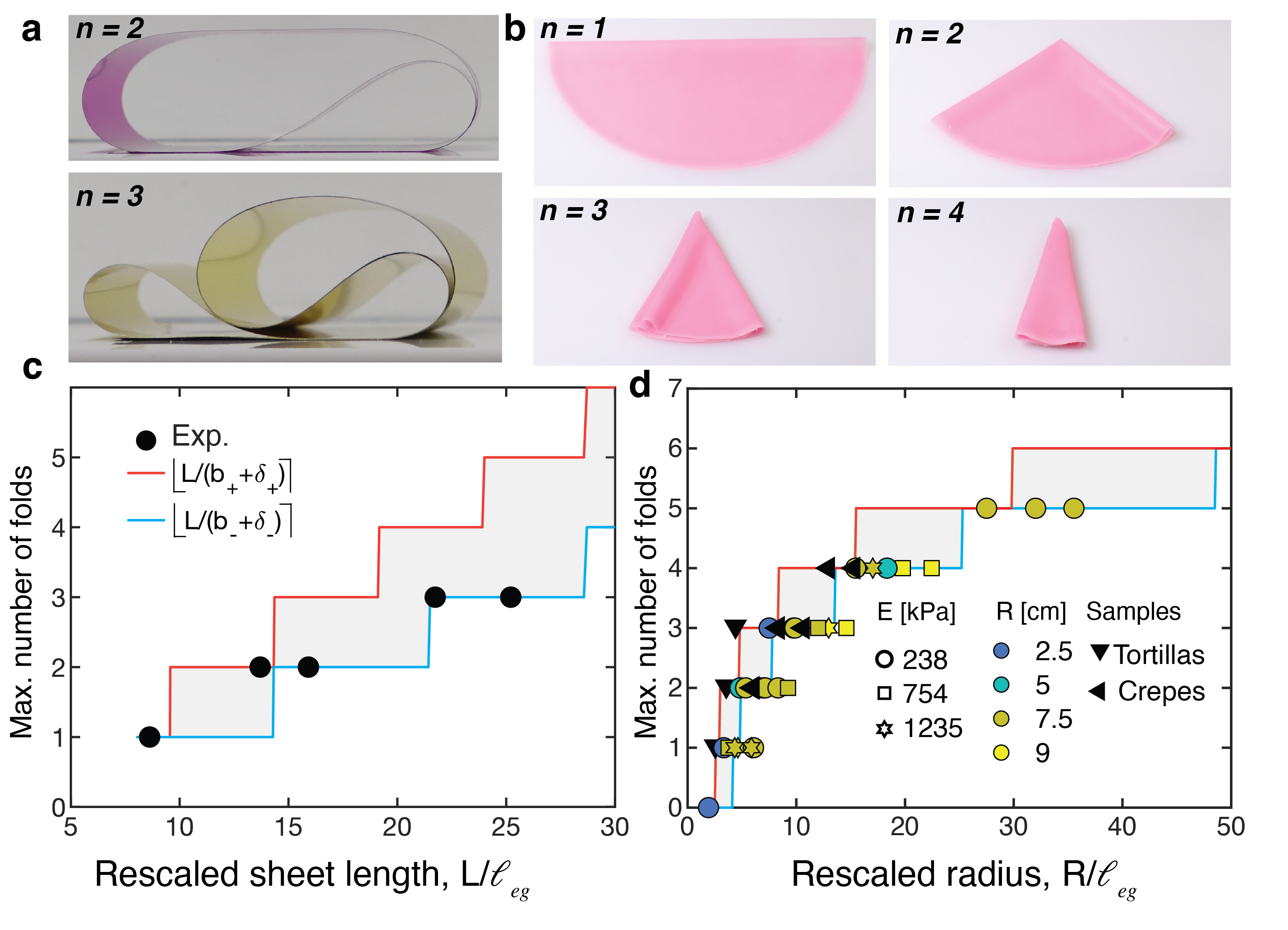}
    \caption{\textbf{Stacking folds:} a) Snapshots of multi-folding states for rectangular sheets with $\ell_{eg}=3.71~\mathrm{cm}$ and $\ell_{eg}=2.4~\mathrm{cm}$, respectively. 
    b) Snapshots of multi-folding states for a circular soft sheet (for $R=7.5~\text{cm}$ and $\ell_{eg}=4.8~\mathrm{cm}$), where the sector angle $\alpha$ is halved at each fold. 
    c)  Maximum number of times a rectangular strip (width $w=6~\mathrm{cm}$ and length $L=60~\text{cm}$) can be folded as a function of the rescaled length $L/\ell_{eg}$. d) Maximum fold count for circular sheets (crêpes) against rescaled radius $R/\ell_{eg}$. 
    In a) and c), contour lines show the geometric prediction; the gray band indicates the admissible range for the fold to be stable.}
    \label{Fig:3}
\end{figure}

In Fig.~\ref{Fig:3}, we report results related to the stability of multi-stack folded configurations for both rectangular and circular sheets made from PET plastic, VPS elastomer, and food samples. For the rectangular case, a ribbon of dimensions $(L \times w) = (60\,\text{cm} \times 6\,\text{cm})$ is folded in half end-to-end, then folded again, and so on, until the configuration becomes unstable and unwraps spontaneously. In Fig.~\ref{Fig:3}a we show the maximum fold count sustained by the sheet as a function of the rescaled length $L/\ell_{eg}$. A smaller ratio yields fewer folds, ranging from 1 fold for the thickest sheets ($h=125\, \mu\text{m}$) to 3 folds for the thinnest ($h=25\, \mu\text{m}$).  

An estimate of the maximum number of folds follows from our previous single-fold analysis and the integration of Eq.~\ref{Eq.elastica}. We argue that each fold spans a length ranging from $\delta_- + b_-= 4.8 \,\ell_{eg}$ and $\delta_+ + b_+ = 7.2 \,\ell_{eg}$. As such, the number of folds is bounded by the nearest integers obtained by dividing the sheet length by these limits.
In Fig.\ref{Fig:3}c, we show that this approach captures all experimental data points, although here the number of folds remains modest. In order to increase the number of folds, we move to circular sheets, as detailed next.

% We arrive at an admissible range of $4.8 \le l_f / \ell_{eg} \le 7.2$ from stability analysis (see Fig.\ref{Fig:2}c); 

%\textcolor{red}{The second photo of (b); $\alpha \sim \pi/2$ seems more than 90 degrees. }
% $R \in [2.5,5,7.5,9]~\text{cm}$
%thicknesses 
We work with circular sheets with radius $R$ (see SI for fabrication details). The folding procedure halves the sector angle $\alpha$ at each step, as illustrated in Fig.~\ref{Fig:3}b. After $n$ folds, the sector angle is $\alpha \approx \pi / 2^{\,n-1}$. In experiments, we vary both the crêpe's radius $R\in[2.5,9]~\text{cm}$ and its elasto-gravity length $\ell_{eg}$, which by~Eq.~\eqref{Eq.elasto-gravity}, depends on the thickness $h\in[0.06,0.97]~\text{mm}$ and the Young's modulus $E\in\{228,884,1101\}~\text{kPa}$ of the sheet. In Fig.~\ref{Fig:3}c, we report the maximum fold number sustained by the sheet. We find that larger radii and smaller values of $\ell_{eg}$ allow for a greater number of folds. In addition to synthetic elastomer sheets, we report data points obtained from industrial food samples, including crêpes and tortillas, as shown in Fig.~\ref{Fig:1}a. Despite their inherent material heterogeneity, these comestible sheets fall within the same stability envelope as the elastomer samples when rescaled by their elasto-gravity lengths (see Fig.~\ref{Fig:3}c), evidencing the robustness of this process.

To rationalize these observations, we use a geometric argument. Each fold halves the sector angle, so that after $n$ folds the available sector area scales as $S \sim \alpha R^{2}$. Stability requires that the area swept by the folded loop, estimated from the characteristic fold size set by $\ell_{eg}$, remains smaller than the available sector area. This criterion yields a closed-form estimate for the maximum fold count $n(R/\ell_{eg})$, which is reported as the stability envelope in Fig.~\ref{Fig:3}c (see SI for derivative details).

\begin{figure}[!h]
    %\centering
    \includegraphics[width=0.85\columnwidth]{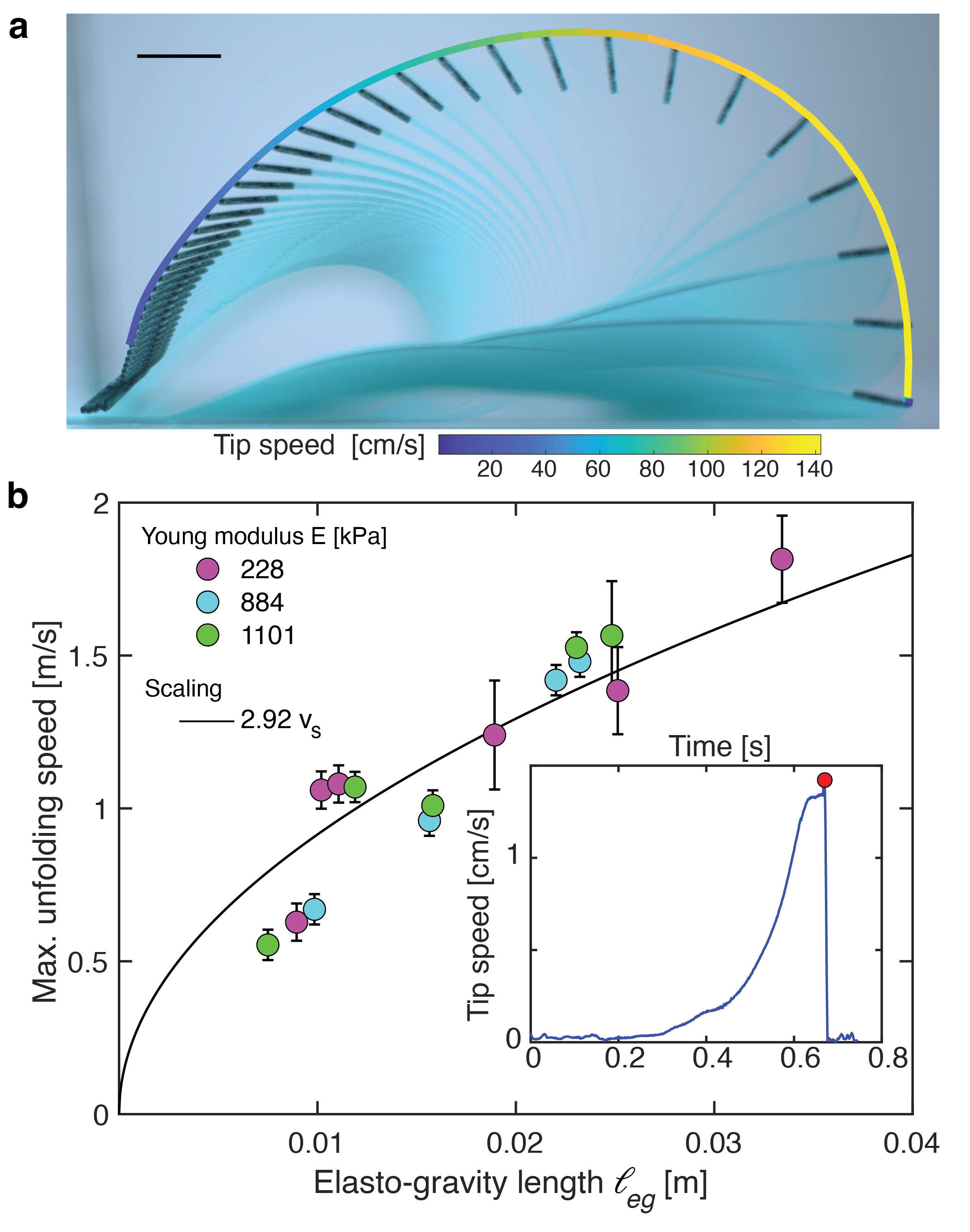}
    \caption{\textbf{Dynamics of unfolding:} a) Chronophotography of the unfolding dynamics for $\ell_{eg}=2.2~\text{cm}$ ($E =\, 884~\text{kPa}$, $h =1.13~\text{mm}$), the solid line reports the tip trajectory and the color modulates the instantaneous speed. Scale bars are $1~\text{cm}$. b) Peak flipping speed as a function of elasto-gravity length [inset: full kinematics of the sample shown in b), red point is the maximal speed].}
    %\Tom{check the value of E, for vps}
    % with the prediction with the corrected PRB model
    \label{Fig:4}
\end{figure}

We note that the folding protocol used for circular sheets introduces a localized geometric singularity at the tip of the folded sector, after the second fold. This singularity appears to be confined to a small region near the apex and affects the overall dynamics across the parameter range explored in this study.

Overall, experiments across all sheet types are well captured by these simple geometric and mechanical arguments. Most data fall closer to the lower bound of the predicted stability range, reflecting the requirement that a stable fold retain a finite flat segment, which increases the material length consumed by each fold. This geometric reasoning parallels Gallivan’s approach for estimating the maximum number of localized creases in paper sheets ~\cite{gallivan2002fold,demaine2015folding,korpal2015say}.

%which debunks the myth that a paper sheet cannot be folded more than eight times. In crease-based folding, each fold is a sharply localized plastic hinge whose spacing is set by yield, allowing many creases to accumulate and without accounting for the material stiffness (purely geometric). In contrast, our folds are smooth elastic loops whose size is governed by the elasto-gravity balance, resulting in fewer folds and a larger material length consumed per fold. This distinction is also functional: whereas plastic creases irreversibly plasticity/damage the sheet, our soft folds remain reversible and preserve the material for repeated un/folding cycles for storage purposes. 

% \subsection*{Dynamics of unfolding}

Finally, we investigate the unwrapping dynamics of a folded elastomer sheet (length $L \in [10,12]$ cm, $\delta \approx \delta_{-} \approx 1.21 \, \ell_{eg}$) and record its motion with a high-speed camera (Chronos 1.4, Kron Technologies -  1069 and 1397 fps). In Fig.~\ref{Fig:4}a, we show a representative chronophotography where the trajectory of the tip is highlighted. We observe a progressive increase in the tip speed, leading to a peak value $v_{\ast} \approx 1.47~\text{m/s}$ attained shortly before the tip impacts the floor (see inset of Fig.~\ref{Fig:4}b). In Fig.~\ref{Fig:4}b, we report the evolution of $v_\ast$ with $\ell_{eg}$ ( $0.2\leq h \leq 1.22$ mm), and observe its increase from $0.5~\text{m/s}$ to nearly $2~\text{m/s}$.

% This peak speed serves as a convenient scalar indicator of the unwrapping dynamics. An extensive study over a range of elasto-gravity lengths $\ell_{eg} \in [0.78-3.39]$~cm was performed by varying the sheet thickness and thickenesses $h \in [0.2-1.22]~\text{mm}$. Fig.~\ref{Fig:4}b shows that the resulting peak speeds clearly increases as a function of $\ell_{eg}$, revealing that sheets with larger elasto-gravity lengths unwrap more rapidly.

We rationalize this dynamics by considering the sum of the bending energy when folded, $E_b \sim B b \ell_{eg}^{-2}$ and of the gravitational potential energy $E_g \sim \rho g h w b^2$, and equate this sum to the kinetic energy in the course of deployment $E_k \sim \rho h w b v_s^2$. Using Eq.~\eqref{Eq.elasto-gravity} and $b \sim \ell_{eg}$ from Fig.~\ref{Fig:2}d, we obtain $v_s\simeq 2.14 \sqrt{g \ell_{eg}}$, a scaling that also dictates the traveling speed of a ruck in a rug~\cite{vella_statics_2009}.  (see SI for derivation details). In Fig.~\ref{Fig:4}b, this scaling is verified with data across all thicknesses and elastomer types, confirming the role of the elasto-gravity in setting a universal time scale for free unfolding. Note that the prefactor differs, as we find $v_s^{exp}\simeq 2.92 \sqrt{g \ell_{eg}}$, a discrepancy that is expected given the crude approximation made in our derivation.

% \section*{Conclusion}
To conclude, we note that a single parameter governs the mechanics of soft folding in heavy elastic sheets: the elasto-gravity length $\ell_{eg}$. 
%This length scale dictates the static geometry of a fold, sets the onset of unfolding, and collapses measurements across materials, sheet dimensions, and configurations. 
Overall, this Letter provides a unified framework for understanding and designing soft folds, from single-loop stability to multi-fold packing and elastically powered actuation. 

\section*{Acknowledgements}
All the authors acknowledge Lauren Dreier, Andrej Košmrlj, Yicong Fu, Crystal Fowler, and Hongsik Kim for helpful discussions; Dominique Marzin for performing the crêpe experiments in Brittany; and Marjorie Batard-Kerviel for the crêpe recipe and cooking advice.
\bibliographystyle{unsrt}  
\bibliography{lit3.bib}

@article{alben2022packing,
  title={Packing of elastic rings with friction},
  author={Alben, Silas},
  journal={Proceedings of the Royal Society A},
  volume={478},
  number={2258},
  pages={20210742},
  year={2022},
  publisher={The Royal Society}
}

@article{jules2019local,
  title={Local mechanical description of an elastic fold},
  author={Jules, Th{\'e}o and Lechenault, Fr{\'e}d{\'e}ric and Adda-Bedia, Mohktar},
  journal={Soft matter},
  volume={15},
  number={7},
  pages={1619--1626},
  year={2019},
  publisher={Royal Society of Chemistry}
}

@book{atanackovic1997stability,
  title={Stability theory of elastic rods},
  author={Atanackovic, Teodor M},
  volume={1},
  year={1997},
  publisher={World Scientific}
}

@article{arriagada2014curling,
  title={Curling and rolling dynamics of naturally curved ribbons},
  author={Arriagada, Octavio Albarr{\'a}n and Massiera, Gladys and Abkarian, Manouk},
  journal={Soft Matter},
  volume={10},
  number={17},
  pages={3055--3065},
  year={2014},
  publisher={Royal Society of Chemistry}
}

@article{batista2020mechanics,
  title={The mechanics of bending a strip of paper},
  author={Batista, Adriano A},
  journal={European Journal of Physics},
  volume={41},
  number={6},
  pages={065009},
  year={2020},
  publisher={IOP Publishing}
}

@article{davidovitch2021rucks,
  title={Rucks and folds: delamination from a flat rigid substrate under uniaxial compression},
  author={Davidovitch, Benny and D{\'e}mery, Vincent},
  journal={The European Physical Journal E},
  volume={44},
  number={2},
  pages={11},
  year={2021},
  publisher={Springer}
}

@article{holmes2019elasticity,
  title={Elasticity and stability of shape-shifting structures},
  author={Holmes, Douglas P},
  journal={Current opinion in colloid \& interface science},
  volume={40},
  pages={118--137},
  year={2019},
  publisher={Elsevier}
}

@article{baroudi_nonlinear_2019,
	title = {Nonlinear dynamics of uniformly loaded \textit{{Elastica}} : {Experimental} and numerical evidence of motion around curled stable equilibrium configurations},
	volume = {99},
	issn = {0044-2267, 1521-4001},
	shorttitle = {Nonlinear dynamics of uniformly loaded \textit{{Elastica}}},
	url = {https://onlinelibrary.wiley.com/doi/10.1002/zamm.201800121},
	doi = {10.1002/zamm.201800121},
	language = {en},
	number = {7},
	urldate = {2025-06-27},
	journal = {ZAMM - Journal of Applied Mathematics and Mechanics / Zeitschrift für Angewandte Mathematik und Mechanik},
	author = {Baroudi, Djebar and Giorgio, Ivan and Battista, Antonio and Turco, Emilio and Igumnov, Leonid A.},
	month = jul,
	year = {2019},
	pages = {e201800121},
}

@article{callan-jones_self-similar_2012,
	title = {Self-{Similar} {Curling} of a {Naturally} {Curved} {Elastica}},
	volume = {108},
	copyright = {http://link.aps.org/licenses/aps-default-license},
	issn = {0031-9007, 1079-7114},
	url = {https://link.aps.org/doi/10.1103/PhysRevLett.108.174302},
	doi = {10.1103/PhysRevLett.108.174302},
	language = {en},
	number = {17},
	urldate = {2025-06-21},
	journal = {Physical Review Letters},
	author = {Callan-Jones, A. C. and Brun, P.-T. and Audoly, B.},
	month = apr,
	year = {2012},
	pages = {174302},
}

@article{deboeuf2024yin,
  title={Yin-Yang spiraling transition of a confined buckled elastic sheet},
  author={Deboeuf, St{\'e}phanie and Proti{\`e}re, Suzie and Katzav, Eytan},
  journal={Physical Review Research},
  volume={6},
  number={1},
  pages={013100},
  year={2024},
  publisher={APS}
}

@inproceedings{demaine2015folding,
  title={Folding a paper strip to minimize thickness},
  author={Demaine, Erik D and Eppstein, David and Hesterberg, Adam and Ito, Hiro and Lubiw, Anna and Uehara, Ryuhei and Uno, Yushi},
  booktitle={International Workshop on Algorithms and Computation},
  pages={113--124},
  year={2015},
  organization={Springer}
}

@book{gallivan2002fold,
  title={How to Fold Paper in Half Twelve Times: An Impossible Challenge Solved and Explained},
  author={Gallivan, Britney C},
  year={2002},
  publisher={Historical Society of Pomona Valley}
}

@article{gomez2017critical,
  title={Critical slowing down in purely elastic ‘snap-through’instabilities},
  author={Gomez, Michael and Moulton, Derek E and Vella, Dominic},
  journal={Nature Physics},
  volume={13},
  number={2},
  pages={142--145},
  year={2017},
  publisher={Nature Publishing Group UK London}
}

@article{korpal2015say,
  title={Say crease! folding paper in half miles please},
  author={Korpal, Gaurish},
  journal={At Right Angles},
  volume={4},
  number={3},
  pages={20--23},
  year={2015},
  publisher={Azim Premji University}
}

@article{lechenault2014mechanical,
  title={Mechanical response of a creased sheet},
  author={Lechenault, Frederic and Thiria, Benjamin and Adda-Bedia, Mokhtar},
  journal={Physical review letters},
  volume={112},
  number={24},
  pages={244301},
  year={2014},
  publisher={APS}
}

@article{brau_wrinkle_2013,
	title = {Wrinkle to fold transition: influence of the substrate response},
	volume = {9},
	issn = {1744-683X, 1744-6848},
	shorttitle = {Wrinkle to fold transition},
	url = {https://xlink.rsc.org/?DOI=c3sm50655j},
	doi = {10.1039/c3sm50655j},
	language = {en},
	number = {34},
	urldate = {2025-07-10},
	journal = {Soft Matter},
	author = {Brau, Fabian and Damman, Pascal and Diamant, Haim and Witten, Thomas A.},
	year = {2013},
	note = {Publisher: Royal Society of Chemistry (RSC)},
	pages = {8177},
}

@article{marzin2025augmented,
  title={Augmented snap-through instability of folded strips},
  author={Marzin, Tom and Venkateswaran, Barath and Baroux, Thomas and Brun, P-T},
  journal={Physical Review Research},
  volume={7},
  number={1},
  pages={013330},
  year={2025},
  publisher={APS}
}

@article{kolinski2009shape,
  title={Shape and motion of a ruck in a rug},
  author={Kolinski, John M and Aussillous, Pascale and Mahadevan, Lakshminarayanan},
  journal={Physical review letters},
  volume={103},
  number={17},
  pages={174302},
  year={2009},
  publisher={APS}
}

@article{paulsen2015optimal,
  title={Optimal wrapping of liquid droplets with ultrathin sheets},
  author={Paulsen, Joseph D and D{\'e}mery, Vincent and Santangelo, Christian D and Russell, Thomas P and Davidovitch, Benny and Menon, Narayanan},
  journal={Nature materials},
  volume={14},
  number={12},
  pages={1206--1209},
  year={2015},
  publisher={Nature Publishing Group UK London}
}

@article{py2007capillary,
  title={Capillary origami: spontaneous wrapping of a droplet with an elastic sheet},
  author={Py, Charlotte and Reverdy, Paul and Doppler, Lionel and Bico, Jos{\'e} and Roman, Benoit and Baroud, Charles N},
  journal={Physical review letters},
  volume={98},
  number={15},
  pages={156103},
  year={2007},
  publisher={APS}
}

@article{sano2023randomly,
  title={Randomly stacked open cylindrical shells as functional mechanical energy absorber},
  author={Sano, Tomohiko G and Hohnadel, Emile and Kawata, Toshiyuki and M{\'e}tivet, Thibaut and Bertails-Descoubes, Florence},
  journal={Communications Materials},
  volume={4},
  number={1},
  pages={59},
  year={2023},
  publisher={Nature Publishing Group UK London}
}

@article{merritt1978pressure,
  title={The pressure curve for a rubber balloon},
  author={Merritt, DR and Weinhaus, F},
  journal={American Journal of Physics},
  volume={46},
  number={10},
  pages={976--977},
  year={1978}
}

@article{taffetani2019limitations,
  title={Limitations of curvature-induced rigidity: How a curved strip buckles under gravity},
  author={Taffetani, Matteo and Box, Finn and Neveu, Arthur and Vella, Dominic},
  journal={Europhysics Letters},
  volume={127},
  number={1},
  pages={14001},
  year={2019},
  publisher={IOP Publishing}
}

@article{tani2024curvy,
  title={Curvy cuts: Programming axisymmetric kirigami shapes},
  author={Tani, Marie and Hong, Joo-Won and Tomizawa, Takako and Lepoivre, {\'E}tienne and Bico, Jos{\'e} and Roman, Beno{\^\i}t},
  journal={Extreme Mechanics Letters},
  volume={71},
  pages={102195},
  year={2024},
  publisher={Elsevier}
}

@article{vella2011wrinkling,
  title={Wrinkling of pressurized elastic shells},
  author={Vella, Dominic and Ajdari, Amin and Vaziri, Ashkan and Boudaoud, Arezki},
  journal={Physical review letters},
  volume={107},
  number={17},
  pages={174301},
  year={2011},
  publisher={APS}
}

@article{vella_statics_2009,
	title = {Statics and {Inertial} {Dynamics} of a {Ruck} in a {Rug}},
	volume = {103},
	copyright = {http://link.aps.org/licenses/aps-default-license},
	issn = {0031-9007, 1079-7114},
	url = {https://link.aps.org/doi/10.1103/PhysRevLett.103.174301},
	doi = {10.1103/physrevlett.103.174301},
	language = {en},
	number = {17},
	urldate = {2025-07-10},
	journal = {Physical Review Letters},
	author = {Vella, Dominic and Boudaoud, Arezki and Adda-Bedia, Mokhtar},
	month = oct,
	year = {2009},
	note = {Publisher: American Physical Society (APS)},
}

@article{wang_critical_1986,
	title = {A critical review of the heavy elastica},
	volume = {28},
	copyright = {https://www.elsevier.com/tdm/userlicense/1.0/},
	issn = {0020-7403},
	url = {https://linkinghub.elsevier.com/retrieve/pii/0020740386900524},
	doi = {10.1016/0020-7403(86)90052-4},
	language = {en},
	number = {8},
	urldate = {2025-07-10},
	journal = {International Journal of Mechanical Sciences},
	author = {Wang, C.Y.},
	month = jan,
	year = {1986},
	note = {Publisher: Elsevier BV},
	pages = {549--559},
}

@article{yoshida2020mechanics,
  title={Mechanics of a snap fit},
  author={Yoshida, Keisuke and Wada, Hirofumi},
  journal={Physical Review Letters},
  volume={125},
  number={19},
  pages={194301},
  year={2020},
  publisher={APS}
}

\end{document}